\documentclass[%
twocolumn,
prl,
]{revtex4}

\usepackage[utf8]{inputenc}
\usepackage[colorlinks=true,linkcolor=red]{hyperref}
\usepackage{url}
\usepackage{amsmath}
\usepackage{amsfonts}
\usepackage{amssymb}
\usepackage{verbatim}
\usepackage{times}
\usepackage{cleveref}
\usepackage{graphicx}% Include figure files
\usepackage{cancel}

\newcommand{\braket}[2]{\langle #1| #2 \rangle}
\newcommand{\bra}[1]{\langle #1|}
\newcommand{\ket}[1]{|#1\rangle}

\def\sech{\operatorname{sech}}
\def\ointshift{\oint_{\Gamma(\pm \frac{i\pi}{4})}^{(\kappa)}}
\def\sLapl{\omega}
\def\onlyone{q}
\def\loschAmpl{\mathcal{G}}
\def\losch{\mathcal{L}}
\def\iniH{H_{\text{\tiny I}}}
\def\finH{H_{\text{\tiny F}}}
\def\finHh{\finH^{\text{h}}}
\def\finHs{\finH^{\text{s}}}
\def\Hhs{H^{\text{h/s}}}
\def\Hh{H^{\text{h}}}
\def\Hs{H^{\text{s}}}
\def\finHhs{\finH^{\text{h/s}}}
\def\fint{t_{\text{\tiny fin}}}
\def\fPhi{\Phi_{\text{\tiny fin}}}
\def\gs{\phi_0}
\def\key{\kappa}
\def\ind{j}
\def\magn{h}

\newcommand{\Ah}[2]{a_{#1 #2}^{\text{\tiny (h)}}}
\newcommand{\As}[2]{a_{#1 #2}^{\text{\tiny (s)}}}
\newcommand{\Ahs}[2]{a_{#1 #2}^{\text{\tiny (h/s)}}}
\newcommand{\titleinfo}{Quenching the magnetic flux in 1d fermionic ring: Loschmidt echo and edge singularity} 
\begin{document}

\title{\titleinfo}% Force line breaks with \\

 \author{Andrea De Luca}%
 \email{andrea.de.luca@lpt.ens.fr}
\affiliation{%
Laboratoire de Physique Th\'eorique de l'ENS \& Institut de Physique Theorique Philippe Meyer\\
 24, rue Lhomond 75005 Paris - France. %\textbackslash\textbackslash
}%

\begin{abstract}
We consider the non-equilibrium dynamics of a system of interacting massless fermions in a ring threaded by a magnetic flux. We focus
on the quench where the flux is initially vanishing and is then turned on.
We show that the definition of the limit of abrupt quench is problematic due to the presence of gauge invariance that has to be taken into account.
We then propose a specific protocol where the dynamics is non-trivial. Employing techniques coming from the Algebraic Bethe-Ansatz, we present an exact formula for the Loschmidt echo valid at all times as a Fredholm determinant at the free fermionic point.
From the analysis of the asymptotic behavior of the Fredholm determinant, we show that the distribution of work done at small energies present an edge singularity whose exponent can be explicitly computed.
Using the correspondence between the edge singularity and the decay of the fidelity at finite-size we propose a general formula for the exponent valid also in the interacting case.
\end{abstract}
\maketitle
%\section{Introduction}
The problem of characterizing the long-time dynamics of a closed quantum system has always attracted a lot of interests because of its fundamental 
implications at the basis of thermodynamics. Novel experiments with cold atoms \cite{bloch2008many} have spurred the attention of the community, since it is now possible to prepare systems where the decoherence effects are sufficiently small and the pure quantum dynamics
can be observed for a large amount of time.

The simplest framework, where this question can be investigated, goes under the name of \textit{quantum quench}. Here a system, described by an Hamiltonian $H(g)$ depending on a parameter $g$, is prepared
in the groundstate of $H(g_i)$ for a given initial value of the parameter $g = g_i$. One supposes that $g$ is then brought to a new value $g_f$ 
so fast that the transient regime where the Hamiltonian is time-dependent can be completely ignored and the evolution is only governed by the final Hamiltonian applied to the initial state. 
This protocol has attracted an enormous effort in the last years but despite this simplification, an exact description of the long-time behavior remains elusive even in one-dimensional and integrable models. Up to now, only few results \cite{calabrese2006time, calabrese2007quantum, brandino2012quench, fioretto2010quantum,  pozsgay2013dynamical,fagotti2013stationary,fagotti2013dynamical, torres2013effects} are known beyond non-interacting Hamiltonians. 
Although the dynamics remains unitary, one can generally expect that for a macroscopic system the long-time expectation value of local quantities may be described 
by a thermal ensemble, since the system is acting as its own thermal bath. Here, the initial state just fixes a finite number of parameters, e.g. the temperature and the chemical potential. 
This picture is however hardly compatible with the experimental results of \cite{kinoshita2006quantum} and it is often supposed to fail in presence of integrability due to the existence of an infinite number of local conserved charges whose values keep memory of many details of the initial state.

However, in most of the cases, it is tacitly assumed that the quantum quench limit, characterized by the abrupt change of the Hamiltonian, is well defined and only few works focused on the study of more general protocols \cite{smacchia2012universal, smacchia2013work} trying
to clarify which features are actually universal, i.e. independent on the specific protocol.
We argue that when the parameter changed during the protocol is associated to a gauge field, the abrupt limit is not well defined
as it breaks the gauge invariance. 
To be more concrete, we consider the XXZ spin chain under periodic boundary conditions
\begin{equation}\label{ham}
\iniH = \sum_{\ind = 1}^N \frac{1}{2}(J_\ind s_\ind^+ s_{\ind+1}^-  + h.c.) + \Delta s_\ind^z s_{\ind+1}^z + \magn s_\ind^z
\end{equation}
where $J_\ind$ and $\Delta$ are real constants and $\magn$ is a uniform magnetic field.
The $s_\ind^\alpha$, for $\alpha = x,y,z$ satisfy the spin $1/2$ algebra and $s_\ind^\pm = s_\ind^x \pm i s_\ind^y$. The periodic boundary conditions impose $s_{N+1}^\alpha = s_1^\alpha$.
Using the Jordan-Wigner transformation (JWT) \cite{lieb1961two}, one can recast the spin $1/2$ into spin-less fermions, where the parameter $\Delta$ controls the interaction strength and $\magn$ becomes the chemical potential. 
We will stick to the spin terminology, except where the fermionic formalism can make the picture clearer.
At time $t=0$, we turn on a magnetic field orthogonal to the ring plane, such that the ring is threaded 
by a magnetic flux $\Phi$. For the sake of clearness, we stress that this magnetic field couples with the orbital angular momentum of the fermionic degrees of freedom and has nothing to do with the $h$ in \eqref{ham}, which instead has the role of fixing the total magnetization.
Practically, this orthogonal field can be realized in the lattice model by the Peierls substitution $J_\ind \to J_\ind e^{i \int_{\ind}^{\ind+1} \vec A \cdot \vec dx}$ where $\vec A$ is the vector potential and the integral is along a path connecting the two lattice sites $\ind, \ind+1$. For simplicity we set $\hbar = 1$.
The gauge-invariant quantity is the additional phase acquired by each fermion while winding around the ring and therefore different gauge choices for $\vec A$ will give rise to equivalent descriptions of the system with $\sum_\ind \arg J_\ind = \Phi$. 
We will consider two extreme examples, that we will address as the homogeneous and singular gauge with the two final Hamiltonians $\finHh$ and $\finHs$ given by the following substitutions in \eqref{ham}: $J_\ind \to e^{\frac{i \Phi}{N}} J_\ind$ for $\finHh$ and 
$J_\ind \to e^{i \delta_{\ind, N} \Phi} J_\ind $ for $\finHs$.
These two Hamiltonians are related by the gauge transformation
\begin{equation}\label{gaugeH}
\finHh= e^{-i \Phi S} \finHs e^{i \Phi S}\;,
\end{equation}
where $S = \sum_\ind \ind s_\ind^z$. 
In the following we will assume that \eqref{ham} is translational invariant, namely $J_\ind = J$ for $\Phi = 0$ and to simplify the notation we set $J=1$. In this case, $\finHs$ (and therefore also $\finHh$) is integrable for arbitrary value of $\Phi$ and $\Delta$ in the framework of Algebraic-Bethe-Ansatz (ABA) \cite{faddeev1996algebraic, korepin1997quantum}
This fact was used in \cite{shastry1990twisted} where the stiffness was employed as a detector of the metal-insulator transition while in \cite{sutherland1990adiabatic} a similar protocol was investigated but in the limit of slow dynamics, where the flux is adiabatically increased.
The realization of artificial gauge fields employing ultra-cold atoms is experimentally feasible and in particular Hamiltonians similar to $\finHhs$ have been realized combining radiofrequency and optical-Raman coupling fields \cite{jimenez2012peierls}. 
We focus on $-1<\Delta<1$ where the system is gapless and we consider the Loschmidt echo $\losch(t)$, a quantity known from different contexts as quantum chaos \cite{jalabert2001environment, karkuszewski2002quantum}, edge singularities \cite{schotte1969tomonaga}, dephasing \cite{quan2006decay}. 
Suppose that the quench protocol is performed from time $t=0$ to $t=\fint$ where the Hamiltonian is time-dependent and is passing from $\iniH$ to $\finH$. 
In this way $\fint$ is the total time duration of the quench procedure. Then, 
$\losch(t) \equiv |\loschAmpl(t)|^2$ where
\begin{equation}\label{LEDef}
\loschAmpl(t) = \bra{\phi(\fint)}e^{i E_0 t}e^{ - i\finH t}\ket{\phi(\fint)} \;,
\end{equation}
$\ket{\phi(\fint)} = \mathcal{U}(\fint) \ket{\phi_0}$ is the evolution at time $\fint$ of the groundstate of the initial Hamiltonian $\iniH$, $E_0$ is the $\iniH$ groundstate energy. 
This quantity can be seen as the generating function of the statistics of work done, i.e. $\loschAmpl(t) = \int dW e^{i W t} P(W)$ \cite{silva2008statistics, smacchia2013work, campisi2011colloquium}, where $P(W)$ is the probability distribution of the work done on the system in the quench protocol.
This expression is gauge invariant, in the sense that it does not change if we transform all quantities with any time-dependent unitary transformation that smoothly becomes the identity at $t=0,\fint$ (see \cite{talkner2007fluctuation} for a discussion).
The Loschmidt echo in the abrupt limit is then defined by taking $\fint \to 0$ where one naively assumes that $\ket{\phi(\fint)} \to \ket{\phi_0}$ in \eqref{LEDef}, which provides the formula for Loschmidt echo in abrupt quenches \cite{silva2008statistics, venuti2011exact}.
However, this is not always correct. To see why, let us assume that the flux smoothly interpolates in a time $\fint$ between $0$ and the final value $\fPhi$, i.e. $\Phi_t = f(t/\fint)\fPhi$, where $f(x)$ is a smooth function with $f(x<0) = 0$ and $f(x>1) = 1$. The corresponding time-dependent Hamiltonians are then given as $\Hhs (t) = \finHhs(\Phi_t)$.
We start focusing on the homogeneous gauge and we take $\Delta = 0$. By JWT, %\footnote{Under periodic boundary conditions for the spin Hamiltonian, the fermionic one presents a non-local boundary term. It can be treated exactly by taking into account the parity of the number of fermions.}
$\finHhs = \sum_{i,j = 1}^N \frac{1}{2} c_i^\dag [\Ahs{i}{j}(\Phi) + h \delta_{ij} ] c_j$, where $\Ah{i}{j} = e^{\frac{i \Phi}{N}} \delta_{j, i+1} + e^{-\frac{i \Phi}{N}} \delta_{i, j+1}$ and
$\As{i}{j} = e^{i \Phi \delta_{i,N}} \delta_{j, i+1} + e^{-i \Phi \delta_{j,N}} \delta_{i, j+1}$. It is then easy to verify that since $\Ah{i}{j}$ is translational invariant,
its eigenvectors are simply given by the Fourier modes and are therefore independent of $\Phi$. It follows that $[\finHh(\Phi_t),\finHh(\Phi_{t'})]=0$, the quench dynamics trivializes and the expression in \eqref{LEDef} reduces to an oscillating phase, such that the work distribution is simply a delta function. The abrupt quench limit can be taken without problems.

Now, we analyze the singular gauge. Clearly, the spectrum of $\As{i}{j}$ is less trivial, since the magnetic flux appears as a local impurity. However, the two operators $\Hhs (t)$ are still related by a unitary operator $U_t = e^{i \Phi_t S}$, since \eqref{gaugeH} holds at any given time $t$ replacing $\Phi$ with $\Phi_t$. This transformation is now time-dependent, so the full gauge transformation for the Hamiltonian operator takes the form
\begin{equation}\label{gaugeTD}
\tilde\Hs (t) \equiv U_t^\dag \Hs(t) U_t + i \frac{d U_t^\dag}{dt}  U_t = \Hh(t) + \frac{f'(t/\fint)}{\fint} S  \;.
\end{equation}
The last term can be interpreted as the lattice discretization of the scalar potential. In fact, for a time-dependent gauge transformation ruled by $\psi(x,t)$, the vector potential $
\vec A$ and the scalar potential $V$ transform as $\vec A \to \vec A + \vec\nabla \psi$ and $V \to V - \partial_t \psi$.
We see now that although $\tilde\Hs(t) = \Hh(t)$ for $ t = 0$ and $\fint$, they differ in between. In particular, 
employing \eqref{LEDef} with the dynamics described by $\tilde\Hs$, we realize that $\ket{\phi(\fint)}$ does not converge to $\ket{\phi_0}$ when $\fint \to 0$, because 
of the last term in \eqref{gaugeTD}, which is divergent for $\fint \to 0$. This argument suggests that the two Loschmidt echoes in the abrupt limit
\begin{equation}\label{LEDAbrupt}
\losch^{\text{h/s}}(t) = |\bra{\phi_0}e^{ - i\finHhs t}\ket{\phi_0}|^2 \;,
\end{equation}
can be different, even though the two final Hamiltonians are gauge-equivalent, because the full time-dependent gauge transformation forbids the naive abrupt limit. 
In particular, $\losch^\text{s}(t)$ comes from the abrupt limit of the dynamics described by $\Hs(t)$ and must be implemented 
by locally changing the hopping coupling $J$ or equivalently employing \eqref{gaugeTD} for sufficiently small $\fint$. Unlike the homogeneous case, 
the result is non-trivial and can be treated for arbitrary $\Delta$ in the framework of ABA. In the thermodynamic limit, it is possible to derive a series expansion at any $t$ which, for $\Delta = 0$, we rewrite in a closed form as a Fredholm determinant.	
In the next, we will refer to singular gauge quantities.

\textit{ABA formulation - } We fix $\cos\zeta  = \Delta$ and we introduce two one-parameter
operators $R(\lambda)$ and $L(\lambda)$, defined on the spaces $\mathbb{C}^2 \otimes \mathbb{C}^2 $, composing a Lax pair,
\begin{equation}\label{rmatrix}
 R = \left(
\begin{array}{llll}
\sinh(\lambda + i\zeta) & 0 & 0 & 0\\
0 & \sinh \lambda & \sinh i \zeta & 0\\
0 & \sinh i\zeta & \sinh\lambda & 0 \\
0 & 0 & 0 & \sinh(\lambda + i\zeta)
 \end{array}\right)	
\end{equation}
and $L(\lambda) = R\left(\lambda - \frac{i\zeta}{2}\right) $. They satisfy the Yang-Baxter equation
\begin{equation}\label{generalYB}
 R_{ab}(\lambda - \mu) L_{na}(\lambda) L_{nb}(\mu) = L_{nb}(\mu) L_{na}(\lambda) R_{ab}(\lambda - \mu)
\end{equation}
where $a,b,n$ label three different copies of $\mathbb{C}^2$ and the equalities should be interpret in the space
$\mathbb{C}^2_a \otimes \mathbb{C}^2_b \otimes \mathbb{C}^2_n$. The quantum monodromy matrix is an operator in $\mathcal{H}\otimes \mathbb{C}^2_a$, where $\mathcal{H} = \mathbb{C}^{2N}$ is the $N$-spin physical space. It is defined as
$T(\lambda) = L_{Na}(\lambda) \ldots L_{1a}(\mu)$ and can be written as a $2\times 2$ matrix of operator-valued entries acting on $\mathcal{H}$
\begin{equation}\label{monodromyABCD}
T(\lambda) = \left( \begin{array}{ll} A(\lambda) & B(\lambda) \\ C(\lambda) &
D(\lambda) \end{array}\right)\;.
\end{equation}
The crucial property is that Eq. \eqref{generalYB} still holds when $L(\lambda)$ is replaced by $T(\lambda)$ and using \eqref{monodromyABCD}, it can be interpreted as a set of commutation relations for $A, B, C, D$. For arbitrary $\key$, 
the transfer matrix $F_\key(\lambda) \equiv A(\lambda) + \key D(\lambda)$ generates a family of commuting operators, i.e. $[F_\key(\lambda), F_\key(\mu)] = 0$. Expanding its logarithm around $i \zeta / 2$, one obtains a set of commuting local operators and in particular 
\begin{equation}\label{hamfromABA}
\finHs = \frac{i  \sin \zeta }{2}\left.\frac{d \ln F_\key(\lambda)}{d \lambda}\right|_{\lambda = i \zeta/2} + \mbox{const.}
\end{equation}
where $\key = e^{i \Phi}$. This shows that for arbitrary $\Phi$ the Hamiltonian belongs to a family of conserved charges in involution and is therefore integrable~\footnote{We refer to the notion of integrability in the Bethe-Ansatz contest. See \cite{caux2011remarks} for a thorough analysis. }. This construction allows expressing 
all the eigenstates in the form $B(\lambda_1) \ldots B(\lambda_M) \ket{0}$, where $\ket{0} = \ket{\uparrow\ldots\uparrow}$ and the roots $\lambda_i$ satisfy the
Bethe-Ansatz equations
\begin{equation}\label{BAE}
N p(\mu_j) - \sum_{k=1}^M \vartheta(\mu_j - \mu_k)  = 2 \pi n_j + \Phi
\end{equation}
for $j = 1,\ldots,M$. Here, the bare momentum is defined as $p(\lambda) = i \log \frac{\sinh (\lambda + \frac{i \zeta}{2})}{\sinh (\lambda - \frac{i \zeta}{2})}$ and the scattering phase as $\vartheta(\lambda) = i \log \frac{\sinh (\lambda + i \zeta)}{\sinh (\lambda - i \zeta)}$.
The $n_j$ must be all distinct and belong to $\mathbb{Z}$ for $M$ odd and to $\mathbb{Z} + \frac 1 2$ for $M$ even. Each choice corresponds to a different solution with energy eigenvalue $E(\lambda_1,\ldots,\lambda_M) = \sum_{j=1}^M \varepsilon(\lambda_j)$ where the single-particle energy
takes the form
\begin{equation}\label{singleparticleenergy}
\varepsilon(\mu) = -\frac{\sin^2 \zeta}{2\sinh(\mu + \frac{i \zeta}{2})\sinh(\mu -
\frac{i \zeta}{2})} \;. 
\end{equation}
The value of $M$ fixes the total magnetization, which is a conserved quantity. For the groundstate in each sector, the $M$ integers must be chosen symmetrically around zero with no gaps and in the thermodynamic limit, one sends $M, N \to \infty$ 
with a constant ratio fixed by $\magn$. The roots becomes denser and denser on a finite interval $[-\Lambda_\magn, \Lambda_\magn]$ of the real line, with $\Lambda_h \stackrel{\magn \to 0}{\longrightarrow} \infty$. They are well described by a density function $\rho(\mu)$ and Eq. \eqref{BAE}
can be converted into a linear integral equation for $\rho(\mu)$.
The advantage of the algebraic formulation is that employing the commutation relations in 
\eqref{generalYB} and the relations in \eqref{BAE}, one can avoid dealing with the explicit expression
of the wave-functions, which can be rather involved. Moreover, the operators $A,B,C,D$ do not depend on $\Phi$, which only affects the roots in \eqref{BAE}. 
We end up with the need of expressing the time-evolution operator appearing in \eqref{LEDef} in terms of $A,B,C,D$. Using the relation \eqref{hamfromABA} and the Trotter formula\cite{trotter1959product}, we get
\begin{equation}\label{trotter}
\loschAmpl(t) = e^{i t E_0} \lim_{L \to \infty} \bra{\gs} F_\key^L\Bigl(\frac{i\zeta}{2}+\frac{t \sin \zeta}{2 L}\Bigr) F_\key^{-L} \Bigl(\frac{i \zeta}{2}\Bigr)
\ket{\gs}
\end{equation} 
The computation of the Loschmidt echo is reduced to the computation of the multiple action of the transfer matrix on the initial groundstate. A similar expression has been used in \cite{kitanine2005dynamical, kitanine2005master, kitanine2002spin}
in order to compute the two-point correlation functions in the groundstate. There, an explicit formula was obtained as a multidimensional contour integral in $M$ variables, where each contour surrounds the $M$ roots. 
In the thermodynamic limit, employing the root density, it can be recast into an infinite series of multiple integrals with an increasing number of variables. Although compact, this expression is hardly computable in general and in \cite{kitanine2009algebraic} a detailed analysis of this formula allowed obtaining the 
large distances expansion of the time-independent correlation function. At the free-fermionic point, i.e. $\Delta = 0$ or $\zeta = -\frac{\pi}{2}$, 
\begin{widetext}
the integral equations involving the root density become exactly solvable and the expansion simplifies to the expression
\begin{equation}\label{loschTLff}
\loschAmpl(t) 
= \sum_{n=0}^{\infty} \frac{\gamma^n}{n!^{2}} \int\limits_{-\Lambda_h}%
^{\Lambda_h} d^n\lambda\sum_{m=0}^n \binom{n}{m} (-\kappa)^{-m} \\
\oint\limits_{\Gamma\{\frac{i \pi}{2}\}}\prod_{j=1}^{m}
\frac{dz_j}{2\pi i}\oint\limits_{\Gamma\{-\frac{i\pi}{2}\}}\prod_{j=m+1}^{n}\frac{dz_j}{2\pi i}
 \left(\det \frac{1}{\sinh (\lambda_j - z_k)}\right)^2
\prod_{b=1}^ne^{it(\varepsilon(\lambda_b)-\varepsilon(_zb))}
\end{equation}
\end{widetext}
where $\gamma = \frac{1 - \key}{2 \pi i}$ and $\Gamma\{\pm\frac{i \pi}{2}\}$ are small contours in the complex plain surrounding $\pm \frac{i \pi}{2}$ and no other singularities.
The energy in \eqref{singleparticleenergy} reduces to $\varepsilon(\lambda) = -\sech2\lambda$.
The interesting point of this expression is that the variables $\lambda_j$/$z_k$ appear in different rows/columns. 
This allows an explicit computation of the integrals over $z_j$ and then the sum over $n$ can be recognized as the series expansion of a Fredholm determinant acting on $[-\Lambda_h, \Lambda_h]$ with $\cosh 2\Lambda_h = 4 /h$ 
(see Suppl. Mat.)%\footnote{See Supplemental Material at [URL] for the details of the derivation of the Fredholm determinant.}
\begin{equation}\label{fredholmexact}
\loschAmpl(t) = \det \left( 1 + \gamma K_t \right)
\end{equation}
where the Kernel is defined as 
\begin{equation}\label{kernelfre}
K_t(\lambda, \mu) \equiv \frac{i  \left[\key+1 +(1 - \key) Q(\lambda,\mu)\right] \sin \Delta_t \varepsilon(\lambda,\mu)}{\kappa \sin(\lambda - \mu)}
\end{equation}
and $\Delta_t \varepsilon(\lambda,\mu) = \frac{(\varepsilon(\lambda) - \varepsilon(\mu)) t}{2}$.
The function $Q(\mu,\nu)$ is expressed using the Fourier transform of the Bessel function on a finite interval $\Omega _{\tau }(\lambda )=\int_0^{\tau } J_0(u) e^{i u \varepsilon(\lambda)}du$, as
\begin{equation}\label{Q}
Q(\lambda, \mu) \equiv \frac{ \Omega _t(\lambda ) \tanh 2 \lambda}{2}   \left(1+i \cot \Delta_t \varepsilon(\lambda,\mu)\right) + (\lambda\leftrightarrow\mu)
 \end{equation}
It is interesting to notice that the dependence on $\key$ is not only in the prefactor $\gamma$ in \eqref{fredholmexact}
but also in the Kernel itself \eqref{kernelfre}. %This can be seen as an effect of the explicit breaking of the time reversal symmetry. 
The universal effects are recovered for large times $t$, corresponding to low-energies in the distribution of work done. For large $t$, the Kernel can be simplified
since $\Omega_{\tau} (\lambda) \stackrel{\tau \to \infty}{\longrightarrow} |\coth 2 \lambda|$ and therefore, changing variable as $\varepsilon(\lambda) = \cos p$, we obtain the compact formula
\begin{equation}\label{asyfre}
\loschAmpl(t) \stackrel{t\gg1}{=} \det \left( 1 + \frac{\gamma  \sin [\frac{t}{2} (\cos p - \cos q) + \theta_{-}(p,q) \Phi	] }{i \sin \frac{1}{2} (p - q)}\right)
\end{equation}
acting on the interval $[-k_F, k_F]$, where $\theta_-(p,q) = \frac 1 2(\theta(p) - \theta(q))$ and $\theta(x)$ is the Heaviside step	 function. 
Notice the additional phase term in the $\sin$ inside the numerator of the Kernel, that gives a different phase factor between the left and right moving particles. 
Similar results were obtained in \cite{colomo1992correlators} as the generating function of the two-point equal-time correlation function in the XX chain. In our case they hold only asymptotically, since we are dealing with the time-dependent case.
This expression is particularly useful because it is possible to extract the leading large $t$ behavior employing general results for Fredholm determinant with a $\sin$-Kernel, that rely on the mapping onto a Riemann-Hilbert problem \cite{kitanine2009riemann}. We get the power-law behavior $\losch(t) \simeq O(t^{-\frac{\Phi^2}{\pi^2}})$. 
\begin{figure}[ht]
\centering
\includegraphics[width=0.8\columnwidth]{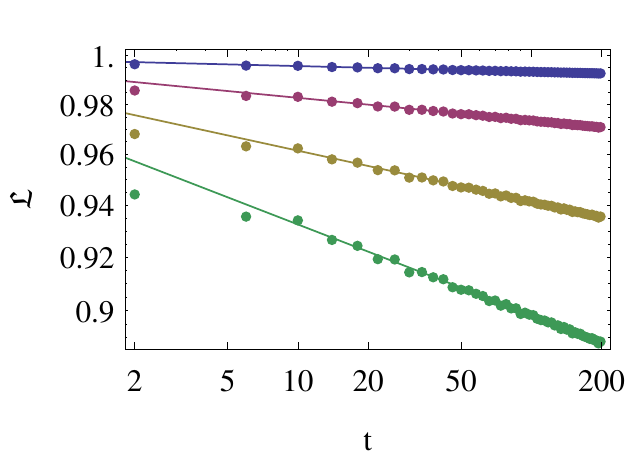}
\caption{The Loschmidt echo computed numerically from \eqref{fredholmexact} with the technique exposed in \cite{bornemann2010numerical} and the corresponding power-law expansion obtained analytically from \cite{kitanine2009riemann} applied to \eqref{asyfre}. The different
curves corresponds to $\Phi = \{0.1, 0.2,0.3,0.4\}$ (top to bottom).
}
\label{plotecho}
\end{figure}
A comparison with \eqref{asyfre} is shown in Fig.~\ref{plotecho}.
This translates into an edge-singularity for the work-distribution $P(W) \simeq (W-\delta E)^{\frac{\Phi^2}{2 \pi^2}-1}$, for $W \gtrsim \delta E$. Here, $\delta E$ is the minimum possible amount of work corresponding to the energy difference between the initial and final groundstates.

This result is consistent with the expectancy of the edge singularity for a local quench in a critical phase. 
With our choice of the gauge, effectively the Hamiltonian is changed only locally. However, 
in principle the change of the magnetic flux is not a local perturbation: only inspecting the quench dynamics, one understands the source of such a local effect.

\textit{Relation to the fidelity - } It has been noticed in \cite{munder2012anderson} that there is a close connection between the edge singularity and the orthogonality catastrophe \cite{anderson1967infrared}.
The change of the Hamiltonian in the quench dynamics induces a change in the single-particle spectrum. This imposes a global rearrangement of the groundstate, such that the initial and final groundstate are indeed orthogonal in the thermodynamic limit.
At finite size, this overlap, dubbed fidelity, shows a power-law decay. For our quench protocol, the overlap can be computed exactly for arbitrary $\Delta$ and $\Phi$ \cite{kitanine2009thermodynamic} and one obtains
\begin{equation}\label{fidelity}
|\braket{\phi_0}{\phi_\Phi}|^2 \equiv O(N^{-\frac{\Phi^2 \mathcal{Z}^2}{2 \pi^2}}) 
\end{equation}
where $\mathcal{Z} = Z(\Lambda_h)$ and $Z(\lambda)$ solves the integral equation
\begin{equation}\label{dressedCharge}
Z(\lambda) + \frac{1}{2\pi} \int_{-\Lambda_h}^{\Lambda_h} \vartheta'(\lambda - \mu) Z(\mu) d \mu = 1 \;. 
\end{equation}
This quantity is naturally interpreted as the intrinsic magnetic moment of the elementary excitations. 
Clearly $Z(\lambda) = 1$ when $\Delta = 0$. 
Consistently with the Anderson interpretation, the exponent in \eqref{fidelity} is equal to
the change in scattering phase shifts at the Fermi surface
divided by $\pi$. In our case, the shift is given by the magnetic flux time the magnetic moment, which is ``dressed'' by the interactions, according to \eqref{dressedCharge}.
As argued in \cite{munder2012anderson}, the large-time exponent in $\losch(t)$ should be twice the exponent of \eqref{fidelity}. This provides a generalization of the exponent of the
edge singularity holding at finite interaction. 
This explains why the edge-singularity exponent in critical local quenches has always a similar structure (compare for instance with \cite{silva2008statistics}). Different physical processes
contribute only to a different phase shift that appears in the same way in the edge-singularity exponent. Here, we see how this result extends to interacting integrable models, by appropriately taking into account the interactions.

\textit{Conclusions - } We investigated the role of gauge invariance in a quantum quench protocol. Final Hamiltonians related by gauge transformation
may practically correspond to different processes. In a specific example, we provided an exact calculation at the free fermionic point for the Loschmidt echo 
valid at arbitrary time as a Fredholm determinant. The result is exact in the lattice and does not rely on the scaling limit. The expression can be expanded for large times, where a power-law behavior emerges
signaling the presence of an edge singularity in the probability distribution of the work done. The comparison of this result with the expression of the fidelity allowed us to extend
the expression of the exponent to the interacting case. Although the Fredholm determinant formula holds only at the free-fermionic point where a more direct approach could have been considered,
the ABA machinery allows for a unified treatment of the interacting case. Even in absence of a closed formula, 
the large-time expansion of the Loschmidt echo is possible along the same lines that lead to the fidelity \cite{kitanine2009thermodynamic, kitanine2011form, kitanine2012form}.
We plan to extend the use of ABA for the computation of Loschmidt echo to all cases of local quenches that conserve
the algebraic structure of the Yang-Baxter equation. 
%%\end{document}

\emph{Acknowledgments - } I would like to thank P. Smacchia for discussions and comments, A. Scardicchio and in particular F. Franchini for carefully reading the manuscript. 
\bibliographystyle{apsrev}
\bibliography{../../../../PHDThesis/References/references,../../../workdistrib}

\begin{thebibliography}{40}
\expandafter\ifx\csname natexlab\endcsname\relax\def\natexlab#1{#1}\fi
\expandafter\ifx\csname bibnamefont\endcsname\relax
  \def\bibnamefont#1{#1}\fi
\expandafter\ifx\csname bibfnamefont\endcsname\relax
  \def\bibfnamefont#1{#1}\fi
\expandafter\ifx\csname citenamefont\endcsname\relax
  \def\citenamefont#1{#1}\fi
\expandafter\ifx\csname url\endcsname\relax
  \def\url#1{\texttt{#1}}\fi
\expandafter\ifx\csname urlprefix\endcsname\relax\def\urlprefix{URL }\fi
\providecommand{\bibinfo}[2]{#2}
\providecommand{\eprint}[2][]{\url{#2}}

\bibitem[{\citenamefont{Bloch et~al.}(2008)\citenamefont{Bloch, Dalibard, and
  Zwerger}}]{bloch2008many}
\bibinfo{author}{\bibfnamefont{I.}~\bibnamefont{Bloch}},
  \bibinfo{author}{\bibfnamefont{J.}~\bibnamefont{Dalibard}}, \bibnamefont{and}
  \bibinfo{author}{\bibfnamefont{W.}~\bibnamefont{Zwerger}},
  \bibinfo{journal}{Rev. Mod. Phys.} \textbf{\bibinfo{volume}{80}},
  \bibinfo{pages}{885} (\bibinfo{year}{2008}).

\bibitem[{\citenamefont{{Calabrese} and {Cardy}}(2006)}]{calabrese2006time}
\bibinfo{author}{\bibfnamefont{P.}~\bibnamefont{{Calabrese}}} \bibnamefont{and}
  \bibinfo{author}{\bibfnamefont{J.}~\bibnamefont{{Cardy}}},
  \bibinfo{journal}{Phys. Rev. Lett.} \textbf{\bibinfo{volume}{96}},
  \bibinfo{eid}{136801} (\bibinfo{year}{2006}),
  \eprint{arXiv:cond-mat/0601225}.

\bibitem[{\citenamefont{Calabrese and Cardy}(2007)}]{calabrese2007quantum}
\bibinfo{author}{\bibfnamefont{P.}~\bibnamefont{Calabrese}} \bibnamefont{and}
  \bibinfo{author}{\bibfnamefont{J.}~\bibnamefont{Cardy}}, \bibinfo{journal}{J.
  Stat. Mech: Theory Exp.} \textbf{\bibinfo{volume}{2007}},
  \bibinfo{pages}{P06008} (\bibinfo{year}{2007}).

\bibitem[{\citenamefont{Brandino et~al.}(2012)\citenamefont{Brandino, De~Luca,
  Konik, and Mussardo}}]{brandino2012quench}
\bibinfo{author}{\bibfnamefont{G.~P.} \bibnamefont{Brandino}},
  \bibinfo{author}{\bibfnamefont{A.}~\bibnamefont{De~Luca}},
  \bibinfo{author}{\bibfnamefont{R.~M.} \bibnamefont{Konik}}, \bibnamefont{and}
  \bibinfo{author}{\bibfnamefont{G.}~\bibnamefont{Mussardo}},
  \bibinfo{journal}{Phys. Rev. B} \textbf{\bibinfo{volume}{85}},
  \bibinfo{pages}{214435} (\bibinfo{year}{2012}).

\bibitem[{\citenamefont{Fioretto and Mussardo}(2010)}]{fioretto2010quantum}
\bibinfo{author}{\bibfnamefont{D.}~\bibnamefont{Fioretto}} \bibnamefont{and}
  \bibinfo{author}{\bibfnamefont{G.}~\bibnamefont{Mussardo}},
  \bibinfo{journal}{New J. Phys.} \textbf{\bibinfo{volume}{12}},
  \bibinfo{pages}{055015} (\bibinfo{year}{2010}).

\bibitem[{\citenamefont{Pozsgay}(2013)}]{pozsgay2013dynamical}
\bibinfo{author}{\bibfnamefont{B.}~\bibnamefont{Pozsgay}},
  \bibinfo{journal}{arXiv preprint arXiv:1308.3087}  (\bibinfo{year}{2013}).

\bibitem[{\citenamefont{Fagotti and Essler}(2013)}]{fagotti2013stationary}
\bibinfo{author}{\bibfnamefont{M.}~\bibnamefont{Fagotti}} \bibnamefont{and}
  \bibinfo{author}{\bibfnamefont{F.~H.} \bibnamefont{Essler}},
  \bibinfo{journal}{arXiv preprint arXiv:1305.0468}  (\bibinfo{year}{2013}).

\bibitem[{\citenamefont{Fagotti}(2013)}]{fagotti2013dynamical}
\bibinfo{author}{\bibfnamefont{M.}~\bibnamefont{Fagotti}},
  \bibinfo{journal}{arXiv preprint arXiv:1308.0277}  (\bibinfo{year}{2013}).

\bibitem[{\citenamefont{Torres-Herrera and Santos}(2013)}]{torres2013effects}
\bibinfo{author}{\bibfnamefont{E.}~\bibnamefont{Torres-Herrera}}
  \bibnamefont{and} \bibinfo{author}{\bibfnamefont{L.~F.}
  \bibnamefont{Santos}}, \bibinfo{journal}{Phys. Rev. E}
  \textbf{\bibinfo{volume}{88}}, \bibinfo{pages}{042121}
  (\bibinfo{year}{2013}).

\bibitem[{\citenamefont{Kinoshita et~al.}(2006)\citenamefont{Kinoshita, Wenger,
  and Weiss}}]{kinoshita2006quantum}
\bibinfo{author}{\bibfnamefont{T.}~\bibnamefont{Kinoshita}},
  \bibinfo{author}{\bibfnamefont{T.}~\bibnamefont{Wenger}}, \bibnamefont{and}
  \bibinfo{author}{\bibfnamefont{D.}~\bibnamefont{Weiss}},
  \bibinfo{journal}{Nature} \textbf{\bibinfo{volume}{440}},
  \bibinfo{pages}{900} (\bibinfo{year}{2006}).

\bibitem[{\citenamefont{Smacchia and Silva}(2012)}]{smacchia2012universal}
\bibinfo{author}{\bibfnamefont{P.}~\bibnamefont{Smacchia}} \bibnamefont{and}
  \bibinfo{author}{\bibfnamefont{A.}~\bibnamefont{Silva}},
  \bibinfo{journal}{Phys. Rev. Lett.} \textbf{\bibinfo{volume}{109}},
  \bibinfo{pages}{037202} (\bibinfo{year}{2012}).

\bibitem[{\citenamefont{Smacchia and Silva}(2013)}]{smacchia2013work}
\bibinfo{author}{\bibfnamefont{P.}~\bibnamefont{Smacchia}} \bibnamefont{and}
  \bibinfo{author}{\bibfnamefont{A.}~\bibnamefont{Silva}},
  \bibinfo{journal}{arXiv preprint arXiv:1305.2822}  (\bibinfo{year}{2013}).

\bibitem[{\citenamefont{Lieb et~al.}(1961)\citenamefont{Lieb, Schultz, and
  Mattis}}]{lieb1961two}
\bibinfo{author}{\bibfnamefont{E.}~\bibnamefont{Lieb}},
  \bibinfo{author}{\bibfnamefont{T.}~\bibnamefont{Schultz}}, \bibnamefont{and}
  \bibinfo{author}{\bibfnamefont{D.}~\bibnamefont{Mattis}},
  \bibinfo{journal}{Ann. Phys.} \textbf{\bibinfo{volume}{16}},
  \bibinfo{pages}{407} (\bibinfo{year}{1961}).

\bibitem[{\citenamefont{Faddeev}(1996)}]{faddeev1996algebraic}
\bibinfo{author}{\bibfnamefont{L.}~\bibnamefont{Faddeev}},
  \bibinfo{journal}{arXiv preprint hep-th/9605187}  (\bibinfo{year}{1996}).

\bibitem[{\citenamefont{Korepin}(1997)}]{korepin1997quantum}
\bibinfo{author}{\bibfnamefont{V.~E.} \bibnamefont{Korepin}},
  \emph{\bibinfo{title}{Quantum inverse scattering method and correlation
  functions}} (\bibinfo{publisher}{Cambridge university press},
  \bibinfo{year}{1997}).

\bibitem[{\citenamefont{Shastry and Sutherland}(1990)}]{shastry1990twisted}
\bibinfo{author}{\bibfnamefont{B.~S.} \bibnamefont{Shastry}} \bibnamefont{and}
  \bibinfo{author}{\bibfnamefont{B.}~\bibnamefont{Sutherland}},
  \bibinfo{journal}{Phys. Rev. Lett.} \textbf{\bibinfo{volume}{65}},
  \bibinfo{pages}{243} (\bibinfo{year}{1990}).

\bibitem[{\citenamefont{Sutherland and
  Sriram~Shastry}(1990)}]{sutherland1990adiabatic}
\bibinfo{author}{\bibfnamefont{B.}~\bibnamefont{Sutherland}} \bibnamefont{and}
  \bibinfo{author}{\bibfnamefont{B.}~\bibnamefont{Sriram~Shastry}},
  \bibinfo{journal}{Phys. Rev. Lett.} \textbf{\bibinfo{volume}{65}},
  \bibinfo{pages}{1833} (\bibinfo{year}{1990}).

\bibitem[{\citenamefont{Jimenez-Garcia
  et~al.}(2012)\citenamefont{Jimenez-Garcia, LeBlanc, Williams, Beeler, Perry,
  and Spielman}}]{jimenez2012peierls}
\bibinfo{author}{\bibfnamefont{K.}~\bibnamefont{Jimenez-Garcia}},
  \bibinfo{author}{\bibfnamefont{L.~J.} \bibnamefont{LeBlanc}},
  \bibinfo{author}{\bibfnamefont{R.~A.} \bibnamefont{Williams}},
  \bibinfo{author}{\bibfnamefont{M.~C.} \bibnamefont{Beeler}},
  \bibinfo{author}{\bibfnamefont{A.~R.} \bibnamefont{Perry}}, \bibnamefont{and}
  \bibinfo{author}{\bibfnamefont{I.~B.} \bibnamefont{Spielman}},
  \bibinfo{journal}{Phys. Rev. Lett.} \textbf{\bibinfo{volume}{108}},
  \bibinfo{pages}{225303} (\bibinfo{year}{2012}).

\bibitem[{\citenamefont{Jalabert and
  Pastawski}(2001)}]{jalabert2001environment}
\bibinfo{author}{\bibfnamefont{R.~A.} \bibnamefont{Jalabert}} \bibnamefont{and}
  \bibinfo{author}{\bibfnamefont{H.~M.} \bibnamefont{Pastawski}},
  \bibinfo{journal}{Phys. Rev. Lett.} \textbf{\bibinfo{volume}{86}},
  \bibinfo{pages}{2490} (\bibinfo{year}{2001}).

\bibitem[{\citenamefont{Karkuszewski et~al.}(2002)\citenamefont{Karkuszewski,
  Jarzynski, and Zurek}}]{karkuszewski2002quantum}
\bibinfo{author}{\bibfnamefont{Z.~P.} \bibnamefont{Karkuszewski}},
  \bibinfo{author}{\bibfnamefont{C.}~\bibnamefont{Jarzynski}},
  \bibnamefont{and} \bibinfo{author}{\bibfnamefont{W.~H.} \bibnamefont{Zurek}},
  \bibinfo{journal}{Phys. Rev. Lett.} \textbf{\bibinfo{volume}{89}},
  \bibinfo{pages}{170405} (\bibinfo{year}{2002}).

\bibitem[{\citenamefont{Schotte and Schotte}(1969)}]{schotte1969tomonaga}
\bibinfo{author}{\bibfnamefont{K.}~\bibnamefont{Schotte}} \bibnamefont{and}
  \bibinfo{author}{\bibfnamefont{U.}~\bibnamefont{Schotte}},
  \bibinfo{journal}{Phys. Rev.} \textbf{\bibinfo{volume}{182}},
  \bibinfo{pages}{479} (\bibinfo{year}{1969}).

\bibitem[{\citenamefont{Quan et~al.}(2006)\citenamefont{Quan, Song, Liu,
  Zanardi, and Sun}}]{quan2006decay}
\bibinfo{author}{\bibfnamefont{H.}~\bibnamefont{Quan}},
  \bibinfo{author}{\bibfnamefont{Z.}~\bibnamefont{Song}},
  \bibinfo{author}{\bibfnamefont{X.}~\bibnamefont{Liu}},
  \bibinfo{author}{\bibfnamefont{P.}~\bibnamefont{Zanardi}}, \bibnamefont{and}
  \bibinfo{author}{\bibfnamefont{C.}~\bibnamefont{Sun}},
  \bibinfo{journal}{Phys. Rev. Lett.} \textbf{\bibinfo{volume}{96}},
  \bibinfo{pages}{140604} (\bibinfo{year}{2006}).

\bibitem[{\citenamefont{Silva}(2008)}]{silva2008statistics}
\bibinfo{author}{\bibfnamefont{A.}~\bibnamefont{Silva}},
  \bibinfo{journal}{Phys. Rev. Lett.} \textbf{\bibinfo{volume}{101}},
  \bibinfo{pages}{120603} (\bibinfo{year}{2008}).

\bibitem[{\citenamefont{Campisi et~al.}(2011)\citenamefont{Campisi, H{\"a}nggi,
  and Talkner}}]{campisi2011colloquium}
\bibinfo{author}{\bibfnamefont{M.}~\bibnamefont{Campisi}},
  \bibinfo{author}{\bibfnamefont{P.}~\bibnamefont{H{\"a}nggi}},
  \bibnamefont{and} \bibinfo{author}{\bibfnamefont{P.}~\bibnamefont{Talkner}},
  \bibinfo{journal}{Rev. Mod. Phys.} \textbf{\bibinfo{volume}{83}},
  \bibinfo{pages}{771} (\bibinfo{year}{2011}).

\bibitem[{\citenamefont{Talkner et~al.}(2007)\citenamefont{Talkner, Lutz, and
  Hanggi}}]{talkner2007fluctuation}
\bibinfo{author}{\bibfnamefont{P.}~\bibnamefont{Talkner}},
  \bibinfo{author}{\bibfnamefont{E.}~\bibnamefont{Lutz}}, \bibnamefont{and}
  \bibinfo{author}{\bibfnamefont{P.}~\bibnamefont{Hanggi}},
  \bibinfo{journal}{arXiv preprint cond-mat/0703189}  (\bibinfo{year}{2007}).

\bibitem[{\citenamefont{Venuti et~al.}(2011)\citenamefont{Venuti, Jacobson,
  Santra, and Zanardi}}]{venuti2011exact}
\bibinfo{author}{\bibfnamefont{L.~C.} \bibnamefont{Venuti}},
  \bibinfo{author}{\bibfnamefont{N.~T.} \bibnamefont{Jacobson}},
  \bibinfo{author}{\bibfnamefont{S.}~\bibnamefont{Santra}}, \bibnamefont{and}
  \bibinfo{author}{\bibfnamefont{P.}~\bibnamefont{Zanardi}},
  \bibinfo{journal}{Phys. Rev. Lett.} \textbf{\bibinfo{volume}{107}},
  \bibinfo{pages}{010403} (\bibinfo{year}{2011}).

\bibitem[{\citenamefont{Trotter}(1959)}]{trotter1959product}
\bibinfo{author}{\bibfnamefont{H.~F.} \bibnamefont{Trotter}},
  \bibinfo{journal}{Proceedings of the American Mathematical Society}
  \textbf{\bibinfo{volume}{10}}, \bibinfo{pages}{545} (\bibinfo{year}{1959}).

\bibitem[{\citenamefont{Kitanine
  et~al.}(2005{\natexlab{a}})\citenamefont{Kitanine, Maillet, Slavnov, and
  Terras}}]{kitanine2005dynamical}
\bibinfo{author}{\bibfnamefont{N.}~\bibnamefont{Kitanine}},
  \bibinfo{author}{\bibfnamefont{J.~M.} \bibnamefont{Maillet}},
  \bibinfo{author}{\bibfnamefont{N.}~\bibnamefont{Slavnov}}, \bibnamefont{and}
  \bibinfo{author}{\bibfnamefont{V.}~\bibnamefont{Terras}},
  \bibinfo{journal}{Nucl. Phys. B} \textbf{\bibinfo{volume}{729}},
  \bibinfo{pages}{558} (\bibinfo{year}{2005}{\natexlab{a}}).

\bibitem[{\citenamefont{Kitanine
  et~al.}(2005{\natexlab{b}})\citenamefont{Kitanine, Maillet, Slavnov, and
  Terras}}]{kitanine2005master}
\bibinfo{author}{\bibfnamefont{N.}~\bibnamefont{Kitanine}},
  \bibinfo{author}{\bibfnamefont{J.~M.} \bibnamefont{Maillet}},
  \bibinfo{author}{\bibfnamefont{N.}~\bibnamefont{Slavnov}}, \bibnamefont{and}
  \bibinfo{author}{\bibfnamefont{V.}~\bibnamefont{Terras}},
  \bibinfo{journal}{Nucl. Phys. B} \textbf{\bibinfo{volume}{712}},
  \bibinfo{pages}{600} (\bibinfo{year}{2005}{\natexlab{b}}).

\bibitem[{\citenamefont{Kitanine et~al.}(2002)\citenamefont{Kitanine, Maillet,
  Slavnov, and Terras}}]{kitanine2002spin}
\bibinfo{author}{\bibfnamefont{N.}~\bibnamefont{Kitanine}},
  \bibinfo{author}{\bibfnamefont{J.}~\bibnamefont{Maillet}},
  \bibinfo{author}{\bibfnamefont{N.}~\bibnamefont{Slavnov}}, \bibnamefont{and}
  \bibinfo{author}{\bibfnamefont{V.}~\bibnamefont{Terras}},
  \bibinfo{journal}{Nucl. Phys. B} \textbf{\bibinfo{volume}{641}},
  \bibinfo{pages}{487} (\bibinfo{year}{2002}).

\bibitem[{\citenamefont{Kitanine
  et~al.}(2009{\natexlab{a}})\citenamefont{Kitanine, Kozlowski, Maillet,
  Slavnov, and Terras}}]{kitanine2009algebraic}
\bibinfo{author}{\bibfnamefont{N.}~\bibnamefont{Kitanine}},
  \bibinfo{author}{\bibfnamefont{K.~K.} \bibnamefont{Kozlowski}},
  \bibinfo{author}{\bibfnamefont{J.~M.} \bibnamefont{Maillet}},
  \bibinfo{author}{\bibfnamefont{N.}~\bibnamefont{Slavnov}}, \bibnamefont{and}
  \bibinfo{author}{\bibfnamefont{V.}~\bibnamefont{Terras}},
  \bibinfo{journal}{J. Stat. Mech: Theory Exp.}
  \textbf{\bibinfo{volume}{2009}}, \bibinfo{pages}{P04003}
  (\bibinfo{year}{2009}{\natexlab{a}}).

\bibitem[{\citenamefont{Colomo et~al.}(1992)\citenamefont{Colomo, Izergin,
  Korepin, and Tognetti}}]{colomo1992correlators}
\bibinfo{author}{\bibfnamefont{F.}~\bibnamefont{Colomo}},
  \bibinfo{author}{\bibfnamefont{A.}~\bibnamefont{Izergin}},
  \bibinfo{author}{\bibfnamefont{V.}~\bibnamefont{Korepin}}, \bibnamefont{and}
  \bibinfo{author}{\bibfnamefont{V.}~\bibnamefont{Tognetti}},
  \bibinfo{journal}{Phys. Lett. A} \textbf{\bibinfo{volume}{169}},
  \bibinfo{pages}{243} (\bibinfo{year}{1992}).

\bibitem[{\citenamefont{Kitanine
  et~al.}(2009{\natexlab{b}})\citenamefont{Kitanine, Kozlowski, Maillet,
  Slavnov, and Terras}}]{kitanine2009riemann}
\bibinfo{author}{\bibfnamefont{N.}~\bibnamefont{Kitanine}},
  \bibinfo{author}{\bibfnamefont{K.~K.} \bibnamefont{Kozlowski}},
  \bibinfo{author}{\bibfnamefont{J.~M.} \bibnamefont{Maillet}},
  \bibinfo{author}{\bibfnamefont{N.}~\bibnamefont{Slavnov}}, \bibnamefont{and}
  \bibinfo{author}{\bibfnamefont{V.}~\bibnamefont{Terras}},
  \bibinfo{journal}{Commun. Math. Phys.} \textbf{\bibinfo{volume}{291}},
  \bibinfo{pages}{691} (\bibinfo{year}{2009}{\natexlab{b}}).

\bibitem[{\citenamefont{Bornemann}(2010)}]{bornemann2010numerical}
\bibinfo{author}{\bibfnamefont{F.}~\bibnamefont{Bornemann}},
  \bibinfo{journal}{Mathematics of Computation} \textbf{\bibinfo{volume}{79}},
  \bibinfo{pages}{871} (\bibinfo{year}{2010}).

\bibitem[{\citenamefont{M{\"u}nder et~al.}(2012)\citenamefont{M{\"u}nder,
  Weichselbaum, Goldstein, Gefen, and von Delft}}]{munder2012anderson}
\bibinfo{author}{\bibfnamefont{W.}~\bibnamefont{M{\"u}nder}},
  \bibinfo{author}{\bibfnamefont{A.}~\bibnamefont{Weichselbaum}},
  \bibinfo{author}{\bibfnamefont{M.}~\bibnamefont{Goldstein}},
  \bibinfo{author}{\bibfnamefont{Y.}~\bibnamefont{Gefen}}, \bibnamefont{and}
  \bibinfo{author}{\bibfnamefont{J.}~\bibnamefont{von Delft}},
  \bibinfo{journal}{Phys. Rev. B} \textbf{\bibinfo{volume}{85}},
  \bibinfo{pages}{235104} (\bibinfo{year}{2012}).

\bibitem[{\citenamefont{Anderson}(1967)}]{anderson1967infrared}
\bibinfo{author}{\bibfnamefont{P.}~\bibnamefont{Anderson}},
  \bibinfo{journal}{Phys. Rev. Lett.} \textbf{\bibinfo{volume}{18}},
  \bibinfo{pages}{1049} (\bibinfo{year}{1967}).

\bibitem[{\citenamefont{Kitanine
  et~al.}(2009{\natexlab{c}})\citenamefont{Kitanine, Kozlowski, Maillet,
  Slavnov, and Terras}}]{kitanine2009thermodynamic}
\bibinfo{author}{\bibfnamefont{N.}~\bibnamefont{Kitanine}},
  \bibinfo{author}{\bibfnamefont{K.~K.} \bibnamefont{Kozlowski}},
  \bibinfo{author}{\bibfnamefont{J.~M.} \bibnamefont{Maillet}},
  \bibinfo{author}{\bibfnamefont{N.}~\bibnamefont{Slavnov}}, \bibnamefont{and}
  \bibinfo{author}{\bibfnamefont{V.}~\bibnamefont{Terras}},
  \bibinfo{journal}{J. Math. Phys.} \textbf{\bibinfo{volume}{50}},
  \bibinfo{pages}{095209} (\bibinfo{year}{2009}{\natexlab{c}}).

\bibitem[{\citenamefont{Kitanine et~al.}(2011)\citenamefont{Kitanine,
  Kozlowski, Maillet, Slavnov, and Terras}}]{kitanine2011form}
\bibinfo{author}{\bibfnamefont{N.}~\bibnamefont{Kitanine}},
  \bibinfo{author}{\bibfnamefont{K.}~\bibnamefont{Kozlowski}},
  \bibinfo{author}{\bibfnamefont{J.}~\bibnamefont{Maillet}},
  \bibinfo{author}{\bibfnamefont{N.}~\bibnamefont{Slavnov}}, \bibnamefont{and}
  \bibinfo{author}{\bibfnamefont{V.}~\bibnamefont{Terras}},
  \bibinfo{journal}{J. Stat. Mech: Theory Exp.}
  \textbf{\bibinfo{volume}{2011}}, \bibinfo{pages}{P12010}
  (\bibinfo{year}{2011}).

\bibitem[{\citenamefont{Kitanine et~al.}(2012)\citenamefont{Kitanine,
  Kozlowski, Maillet, Slavnov, and Terras}}]{kitanine2012form}
\bibinfo{author}{\bibfnamefont{N.}~\bibnamefont{Kitanine}},
  \bibinfo{author}{\bibfnamefont{K.}~\bibnamefont{Kozlowski}},
  \bibinfo{author}{\bibfnamefont{J.}~\bibnamefont{Maillet}},
  \bibinfo{author}{\bibfnamefont{N.}~\bibnamefont{Slavnov}}, \bibnamefont{and}
  \bibinfo{author}{\bibfnamefont{V.}~\bibnamefont{Terras}},
  \bibinfo{journal}{J. Stat. Mech: Theory Exp.}
  \textbf{\bibinfo{volume}{2012}}, \bibinfo{pages}{P09001}
  (\bibinfo{year}{2012}).

\bibitem[{\citenamefont{Caux and Mossel}(2011)}]{caux2011remarks}
\bibinfo{author}{\bibfnamefont{J.-S.} \bibnamefont{Caux}} \bibnamefont{and}
  \bibinfo{author}{\bibfnamefont{J.}~\bibnamefont{Mossel}},
  \bibinfo{journal}{J. Stat. Mech: Theory Exp.}
  \textbf{\bibinfo{volume}{2011}}, \bibinfo{pages}{P02023}
  (\bibinfo{year}{2011}).

\end{thebibliography}

\clearpage
\newpage

\clearpage 
\setcounter{equation}{0}%
\setcounter{figure}{0}%
\setcounter{table}{0}%
\renewcommand{\thetable}{S\arabic{table}}
\renewcommand{\theequation}{S\arabic{equation}}
\renewcommand{\thefigure}{S\arabic{figure}}

\onecolumngrid

\begin{center}
{\Large Supplementary Material for EPAPS \\ 
\titleinfo
}
\end{center}
\section{Derivation of the Fredholm determinant}
In \cite{kitanine2005dynamical}, it is derived the expression 
%\begin{widetext}
\begin{multline} % 6.16 of \cite{kitanine2005dynamical}
\label{loschTL}
\loschAmpl(t) 
=
\sum_{n=0}^{\infty}\frac{1}{(n!)^2}\int\limits_{-\Lambda_h}%
^{\Lambda_h} d^n\lambda
\oint\limits_{\Gamma\{\pm\frac{i\zeta}2\}}\prod_{j=1}^{n}
\frac{dz_j}{2\pi i}\cdot \prod_{a,b=1}^n\frac{
\sinh(\lambda_a-z_b+i \zeta)\sinh(z_b-\lambda_a+i \zeta)}
{\sinh(\lambda_a-\lambda_b+i \zeta)\sinh(z_a-z_b+i \zeta)}
          \\
\times
\prod_{b=1}^ne^{it(\varepsilon(\lambda_b)-\varepsilon(z_b))}\;
\det_n\tilde M_{\kappa}(\{\lambda\},\{z\})
\cdot\det_n[{\cal R}^{\kappa}_n(\lambda_j,z_k|\{\lambda\},\{z\})]
\end{multline}
where 
\begin{equation}\label{GFtiMjk}
(\tilde M_\kappa)_{jk}(\{\lambda_{\alpha_+}\},\{z\})
=t(z_k,\lambda_j)+\kappa t(\lambda_j,z_k)
\prod_{a\in\alpha_+}\frac{\sinh(\lambda_a-\lambda_j+i\zeta)}
{\sinh(\lambda_j-\lambda_a+i\zeta)}\cdot
\prod_{a=1}^n\frac{\sinh(\lambda_j-z_a+i\zeta)}
{\sinh(z_a-\lambda_j+i\zeta)},
\end{equation}
\begin{equation}\label{cal-R}
{\cal R}^{\kappa}_n(\lambda,z|
\{\lambda\},\{z\})=\left\{
\begin{array}{lr}
\rho(\lambda,z),& z\sim i\zeta/2;\\
-\kappa^{-1}\rho(\lambda,z+i\zeta)
\prod\limits_{b=1}^{n}\frac{\sinh(z-\lambda_b+i\zeta)
\sinh(z_b-z+i\zeta)}{\sinh(\lambda_b-z+i\zeta)\sinh(z-z_b+i\zeta)},
& z\sim-i\zeta/2.
\end{array}\right.
\end{equation}
where $t(\lambda, \mu)$ and $\rho(\lambda, z)$ are defined by
\begin{equation}\label{trhodef}
t(\lambda, \mu) = \frac{\sinh i \zeta}{\sinh(\lambda - \mu) \sinh ( \lambda - \mu + i\zeta)}, \qquad -2 \pi i \rho(\lambda, z) + \int_{-\Lambda_h}^{\Lambda_h} \vartheta'(\lambda- \mu) \rho(\mu,z) d\mu = t(\lambda, z) \;.
\end{equation}
%\end{widetext}
Eq. \eqref{loschTL} is greatly simplified at the free fermion point, where
$\Delta \to 0$, or $\zeta = -\frac{\pi}{2}$. In fact, it is possible 
to provide an explicit expression for the inhomogeneous spectral density
\begin{equation}\label{inhospectraldensity}
\rho(\lambda, z) = \frac{i}{\pi \sinh 2 (\lambda - z)}
\end{equation}
and from \eqref{cal-R}, we derive
\begin{equation}\label{Rff}
{\cal R}^{\kappa}_n(\lambda,z|
\{\lambda\},\{z\})=\left\{
\begin{array}{lr}
\frac{i}{\pi \sinh 2 (\lambda - z)} ,& z\sim - \frac{i\pi}{4};\\
\frac{i \kappa^{-1}}{\pi \sinh 2 (\lambda - z)},
& z\sim\frac{i\pi}{4}.
\end{array}\right. \;,
\end{equation}
while the matrix $\tilde M$ simplifies to
\begin{equation}\label{tildeMff}
(\tilde M_\kappa)_{jk} (\{\lambda\} | \{ z \}) = \frac{2 ( \kappa - 1)}{ \sinh 2 (\lambda_j - z_k)} \;. 
\end{equation}
Moreover, since each integral is symmetric under the permutations of ${z_1, \ldots, z_n}$, we can rewrite
\begin{equation}\label{combinatorialExp}
\oint\limits_{\Gamma\{\pm\frac{i\zeta}{2}\}}\prod_{j=1}^{n}
\frac{dz_j}{2\pi i} = \sum_{m=0}^n \binom{n}{m} \oint\limits_{\Gamma\{-\frac{i\zeta}2\}}\prod_{j=1}^{m}
\frac{dz_j}{2\pi i}\oint\limits_{\Gamma\{\frac{i\zeta}2\}}\prod_{j=m+1}^{n}
\frac{dz_j}{2\pi i} \;.
\end{equation}
%\begin{widetext}
By the explicit expression of Cauchy determinants in \eqref{gencauchy}, we see that
\begin{equation}\label{cauchyHalves}
\left(\det \frac{1}{\sinh 2 (\lambda_j - z_k)}\right)^2  \prod_{a,b=1}^n \frac{\cosh(\lambda_a - z_b)^2}{\cosh(\lambda_a - \lambda_b)\cosh(z_a - z_b)} = 
2^{-2n} \left(\det \frac{1}{\sinh (\lambda_j - z_k)}\right)^2
\end{equation}
and employing \cref{inhospectraldensity,Rff,combinatorialExp,tildeMff} in \eqref{loschTL}, we get \eqref{loschTLff}.
%\end{widetext}
Using the identity in \eqref{detIntegral} 
this expression can be recast as
\begin{equation}\label{fredholm}
\loschAmpl(t) = \det \left( 1 + \gamma K_t \right)
\end{equation}
and the Kernel is defined as
\begin{equation}\label{kernelfreOrig}
K_t(\lambda, \lambda') \equiv \ointshift \frac{dz}{2 \pi i} \frac{\exp\bigl[\frac{it}{2} (\varepsilon(\lambda) + \varepsilon(\lambda') - 2 \varepsilon(z)) \bigr]}{\sinh(z-\lambda)\sinh(z-\lambda')} \;. 
\end{equation}
where in order to simplify the notation, we set
\begin{equation}\label{intshift}
\ointshift = \oint_{\Gamma(-\frac{i\pi}{4})} + \kappa^{-1} \oint_{\Gamma(\frac{i\pi}{4})} \;.  
\end{equation}
The main difficulty in the computation of the Kernel is related to the fact that the integrand has an essential singularity around the points $\pm \frac{i \pi}{4}$ inside the contours. 
In order to deal with it, we consider the time Laplace transform, defined as the integral
%\begin{widetext}
\begin{equation}\label{kernelLapl}
  \tilde K_\sLapl(\lambda, \lambda') \equiv \int_0^\infty dt K_t (\lambda, \lambda') \exp\bigl(-\sLapl t  - \frac{i t}{2} (\varepsilon(\lambda) + \varepsilon(\lambda'))\bigr) = 
 \ointshift \frac{dz}{2 \pi i\sinh(z-\lambda)\sinh(z-\lambda') (\sLapl + i \varepsilon(z))} 
\end{equation}
In this way the essential singularities inside the exponential are converted into simple poles. Notice that convergence of the Laplace transform 
requires $\sLapl$ to be large enough that the poles of the last term in the denominator lie inside the integration domain. 
%Employing that at the free-fermion point $E(\lambda) = - \sech 2\lambda$, the integral can be rewritten as
%$$ \tilde K_\sLapl(\lambda, \lambda')  = \ointshift \frac{dz \cosh2z (\cosh(\lambda + \lambda') \cosh2z  + \sinh(\lambda + \lambda') \sinh2 z - \cosh(\lambda - \lambda'))}{\pi i \omega (\cosh 2z - \cosh 2 \lambda) (\cosh 2z - \cosh 2 \lambda') (\cosh 2 z - \frac{i}{\omega})} \; .$$
%\end{widetext}
It can be computed taking the residues at these poles and we get
\begin{equation}\label{kernelLaplComputed}
\tilde K_\sLapl(\lambda, \lambda') = \frac{\tilde \onlyone_\omega (\lambda) - \tilde \onlyone_\omega(\lambda')}{\sinh(\lambda - \lambda')}
%\tilde K_\sLapl^-(\lambda, \lambda') + \kappa^{-1}\tilde K_\sLapl^+(\lambda, \lambda') 
\end{equation}
%where we set
%\begin{equation}\label{kernelpm}
% \tilde K_\sLapl^{\pm}(\lambda, \lambda')  = \frac{4 (\omega  \cosh (\lambda -\lambda')+4 i \cosh (\lambda +\lambda'))}{\sqrt{\omega^2+16} (\omega  \cosh 2 \lambda +4 i) (\omega  \cosh 2 \lambda'+4 i)} -\frac{4 i \sinh (\lambda +\lambda')}{(\omega  \cosh 2 \lambda +4 i) (\omega  \cosh 2 \lambda'+4 i)}
%\end{equation}
where we set
\begin{equation}\label{onlyone}
\tilde\onlyone_\omega (\lambda) =\frac{(\key -1) \sinh 2 \lambda -(\key +1) \sqrt{\omega ^2+1} \cosh 2 \lambda }{2 \key  (\omega 
   \cosh 2 \lambda -i)  \sqrt{\omega ^2+1}} \;.
\end{equation}
By taking the inverse Laplace transform, we recover the Kernel in \eqref{kernelfre}. 
%Employing \eqref{kernelLaplComputed}, it is expressed with
%\begin{equation}\label{onlyoneinverse}
%\onlyone_{t}(\lambda )=\frac{e^{-i t \varepsilon(\lambda)} \left((\kappa -1) \tanh (2 \lambda ) \Omega _t(\lambda )-1 -\kappa \right)}{2 \kappa} \;.
%\end{equation}
\section{Generalized Cauchy Matrix}
\label{appendixCauchy}
Given two sequences, we can define a Cauchy-Matrix as 
$$ M_{ij} = \frac{1}{x_i-y_j} $$
Then one can prove that the determinant of this matrix has an explicit
expression 
$$ \det M = \frac{\prod_{i<j} (x_i - x_j) (y_j - y_i)}{\prod_{i,j=1}^n (x_i -
y_j)} $$
It is interesting to notice that the previous approach can be generalized to 
$ M_{ij} = (\sinh\alpha (x_i-y_j))^{-1}$
giving the result
\begin{equation}\label{gencauchy}
 \det M = \frac{\prod_{i<j} \sinh\alpha(x_i - x_j) \sinh\alpha(y_j -
y_i)}{\prod_{i,j=1}^n \sinh\alpha(x_i -
y_j)} 
\end{equation}
\section{Determinant integration}
We will prove here the following equality
\begin{equation}\label{detIntegral}
\int \mathcal{D}\mathbf{z} \left[\det
\left(
 g_j(z_i) \right)\right]^2 = N! \det G_{ij} 
\end{equation}
where the measure $\mathcal{D}\mathbf{z} = \prod_{n=1}^N \mu(z_i) d z_i$ and $\mu(z)$ is a measure on the complex plane. We set
\begin{equation}\label{matrixElements}
G_{ij} = \int dz g_i(z) g_j(z) \mu(z) \; .
\end{equation}
It is actually useful to prove the more general expression
%\begin{widetext}
\begin{equation}\label{cauchyProductGeneral}
\int \mathcal{D}\mathbf{z}
\det
\left(
 g_{\sigma(j)}(z_i) \right) \det
\left(
 g_{\eta(j)}(z_i) \right)
 = M! \det
G_{\sigma(i), \eta(j)}
\end{equation}
%\end{widetext}
%\oint_{\mathcal{C}(\{ \lambda\})} \prod_{n=1}^M \frac{f(z_n)	 dz_n}{2 \pi i}
%\det%
%\left(
%\frac{1}{z_j - \lambda_{\sigma(k)}} \right) \det
%\left(
%\frac{1}{z_j - \lambda_{\eta(k)}} \right) = M! \det
%\left(\frac{f_{\sigma(j)}-f_{\eta(k)}}{\lambda_{\sigma(j)} - \lambda_{\eta(k)}}
%\right)
%\end{equation}
where $\sigma, \eta$ are two injective map of $\{1,\ldots, M\}$ in $\{1,\ldots,
N\}$ and $k$ runs from $1$ to $m$. In order to prove it, we notice that each row of the matrix just depend on one
variable $z_i$ and this allows to factorize the multiple integral. 
We observe that \eqref{cauchyProductGeneral} reduces to \eqref{detIntegral}
when $M = N$. We will prove it by induction over $M$. 
\begin{itemize}
 \item for $M = 1$ the integral reduces to one variable and the equality holds.
 \item now we assume the identity for $M-1$ and we prove it for $M$. We expand
the determinants with respect to the first row using the
Laplace formula
$$\det
\left(
g_{\sigma(k)}(z_j)
\right) = \sum_{n=1}^M 
(-1)^{n+1} g_{\sigma(n)}(z_1)	
\det \left( g_{\sigma(k)}(z_j)
\right)_{\substack{j\neq1\\k\neq n}} $$
and a similar equality when $\sigma$ is replaced by $\eta$. Inserting these
expansions in \eqref{cauchyProductGeneral}, the integral over $z_1$ can be readily performed, 
using the fact that the determinants do not involve $z_1$ anymore. So one gets
%\begin{widetext}
$$ (M-1)!\sum_{l,r=1}^M (-1)^{l+r} G_{\sigma(l), \eta(j)} \int  \left[\prod_{n=2}^M \mu(z_n) dz_n \right]
\det \left( g_{\sigma(k)}(z_j)\right)_{\substack{j\neq1\\ k\neq l}} \det \left( g_{\eta(k)}(z_j)
\right)_{\substack{j\neq1\\k\neq r}}  $$
%\end{widetext}
and now the integral involves only $M-1$ variables, so can be computed using the
inductive hypothesis
$$ (M-1)!\sum_{l=1}^M \sum_{r=1}^M (-1)^{l+r} G_{\sigma(l), \eta(j)} \det
\left(G_{\sigma(i), \eta(j)}\right)_{\substack{j\neq l\\k\neq r}} \; .$$
Now the inner sum corresponds to the expansion of the rhs of
\eqref{cauchyProductGeneral} with respect to the row $l$. So, all the terms are
equal and the sum over $l$ just gives a factor $M$ in front of everything
completing the $M!$.
\end{itemize}
\end{document}